\documentclass[pdflatex,sn-nature, iicol]{sn-jnl}
\usepackage{graphicx}%
\usepackage{multirow}%
\usepackage{amsmath,amssymb,amsfonts}%
\usepackage{xcolor}%
\usepackage{textcomp}%
\usepackage[
        range-phrase=--,
        range-units=single,
        retain-explicit-plus=true,
        retain-unity-mantissa=false,
    ]{siunitx}
\usepackage[capitalise]{cleveref}
\newcounter{extfig}
\renewcommand{\theextfig}{\arabic{extfig}}
\crefname{extfig}{Extended Data Fig.}{Extended Data Figs.}
\Crefname{extfig}{Extended Data Fig.}{Extended Data Figs.}
\newcommand{\extfigurecaption}[1]{%
\refstepcounter{extfig}%
\caption*{\textbf{Extended Data Fig. \theextfig} #1}}
\usepackage{caption}
\usepackage{booktabs}   
\usepackage{makecell}
\makeatletter
\renewcommand{\@maketitle}{\null%
    \if@remarkboxon\vbox to 0pt{\vspace*{-78pt}\hspace*{-18pt}\FMremark}\else\vskip21pt\fi
    \hsize\textwidth\parindent0pt
    {\hbox to \textwidth{{\Artcatfont\ArtType\hfill}\par}}
    \ifx\@title\empty\else%
        \removelastskip\vskip20pt\nointerlineskip%
        {\Titlefont\@title\par}
    \fi%
    \ifx\@subtitle\empty\else%
        \vskip9pt%
        {{\SubTitlefont\@subtitle\par}}
    \fi%
    \ifnum\aucount>0
        \global\punctcount\aucount%
        \vskip20pt%
        \artauthors\par
        {\vskip7pt\addressfont\auaddress\par
	 \removelastskip\vskip24pt%
	\ifnum\emailcnt>0\relax%
           \ifx\corrauthemail\@empty\else{\ifnum\aucount>1*\fi}%
	   Corresponding author(s). E-mail(s): \corrauthemail\par\fi%
	   \ifx\authemail\@empty\else Contributing authors:\ \authemail\fi%
        \fi%
        \ifequalcont{\par$^{\dagger}$\@equalconttext\par}\fi%
	 \removelastskip\vskip24pt%
        \ifpresentaddress{\par\@presentaddresstext\par}\fi%
	}
     \fi%
     {\printabstract\par}%
     {\printkeywords\par}%
     \ifx\@pacs\empty\else%
       \loop\ifnum\PacsCount>0%
          \csname\romannumeral\PacsTmpCnt StorePacsTxt\endcsname\par%
          \StepDownCounter{\PacsCount}%
          \StepUpCounter{\PacsTmpCnt}%
       \repeat%
    \fi%
    \removelastskip\vskip36pt\vskip0pt}%
\makeatother

\begin{document}

\title[Article Title]{Few-cycle electro-optic light on thin-film lithium niobate}

\author[1,2]{\fnm{Xinyi} \sur{Ren}}
\equalcont{These authors contributed equally to this work.}
\author[1,2]{\fnm{Chun-Ho} \sur{Lee}}
\equalcont{These authors contributed equally to this work.}

\author[1]{\fnm{Ian} \sur{Christen}}
\author[1,2]{\fnm{Clayton} \sur{Cheung}}
\author[1,2]{\fnm{Reshma} \sur{Kopparapu}}
\author[1,2,3]{\fnm{Yue} \sur{Yu}}
\author[1,2]{\fnm{Lian} \sur{Zhou}}
\author[1,2]{\fnm{Zaijun} \sur{Chen}}
\author*[1,2,3]{\fnm{Mengjie} \sur{Yu}}\email{mengjie.yu@berkeley.edu}

\affil[1]{\orgdiv{Department of Electrical Engineering and Computer Sciences}, \orgname{University of California}, \orgaddress{\city{Berkeley}, 
\state{CA}, \postcode{94720}, \country{USA}}}

\affil[2]{\orgdiv{Ming Hsieh Department of Electrical and Computer Engineering}, \orgname{University of Southern California}, \orgaddress{\city{Los Angeles}, \state{CA}, \postcode{90089}, \country{USA}}}

\affil[3]{\orgdiv{Materials Sciences Division}, \orgname{Lawrence Berkeley National Laboratory}, \orgaddress{\city{Berkeley}, 
\state{CA}, \postcode{94720}, \country{USA}}}

\abstract{
\textbf{
The twin fields of ultrafast optics and nonlinear photonics enable numerous applications ranging from attosecond science~\cite{Popmintchev_HHG_2012,Hu_attosecond_lasers_2026} and ultrafast electronics~\cite{Rybka_nanotunnelling_2016} to molecular spectroscopy~\cite{Wen_dispersive_FT_2026,Picque_freqcomb_spectroscopy_2019}, nonlinear optics~\cite{Wu_UVcomb_2024,Fortier2019FrequencyComb,Krogen_MIR_singlecycle_2017}, quantum nanophotonics~\cite{Sosnicki_picosecond_nanosecond_2023} and precision metrology~\cite{Na_TOF_2020}. However, bringing these capabilities---including  ultrashort pulse generation, dispersion control, and strong nonlinear interactions---together within a scalable photonic integrated platform requires exceptional performance and cooperation between components while at the same time preserving sufficient optical power across the circuit: a level of system integration that had not previously been achieved. Here we demonstrate an integrated multi-functional nonlinear photonic system that transforms continuous-wave (CW) light into high-peak-power femtosecond pulses and directly harnesses them for pulse-driven nonlinear optics on thin-film lithium niobate (TFLN). Microwave-driven electro-optic (EO) broadening followed by integrated dispersive compression generates 230-fs Fourier-transform-limited pulses with energies up to 3.3 pJ at a microwave rate of 30.7 GHz, representing orders of magnitude higher pulse energy than previous integrated pulse synthesis at comparable repetition rates~\cite{Yu_femtosecondLN_2022}. The pulses drive two complementary nonlinear architectures. In a 0.3-meter dispersion-engineered TFLN waveguide, soliton dynamics nonlinearly compress the pulses to 35 fs (6.7 optical cycles), accompanied by coherent spectral broadening exceeding 330 nm. In a fully-monolithic architecture, EO synthesis, dispersive compression and a high-$Q$ nonlinear resonator are integrated on a single TFLN chip, enabling resonantly-enhanced pulse pumping and coherent spectral broadening at pulse energies as low as 400 fJ. By unifying microwave-controlled pulse synthesis and pulse-driven nonlinear interactions, our work establishes a direct path from CW excitation to few-cycle nonlinear optics on chip. More broadly, it establishes TFLN as a scalable nonlinear platform for compact, electronically-synchronized ultrafast photonics, with opportunities spanning microwave photonics~\cite{Ye_Brillouin_TFLN_2025}, optical frequency synthesis and metrology, and mid-infrared and terahertz generation~\cite{Didier_MIR_TFLN_2026,Herter_THz_TFLN_2023}.
}
}

{\dimen0=\textheight\addtolength{\textheight}{2cm}\maketitle\global\textheight=\dimen0}
\,
\newpage
\newpage
\section{Introduction}\label{sec1}

\begin{figure*}[t]   
    \centering
    \includegraphics[width=\textwidth]
    {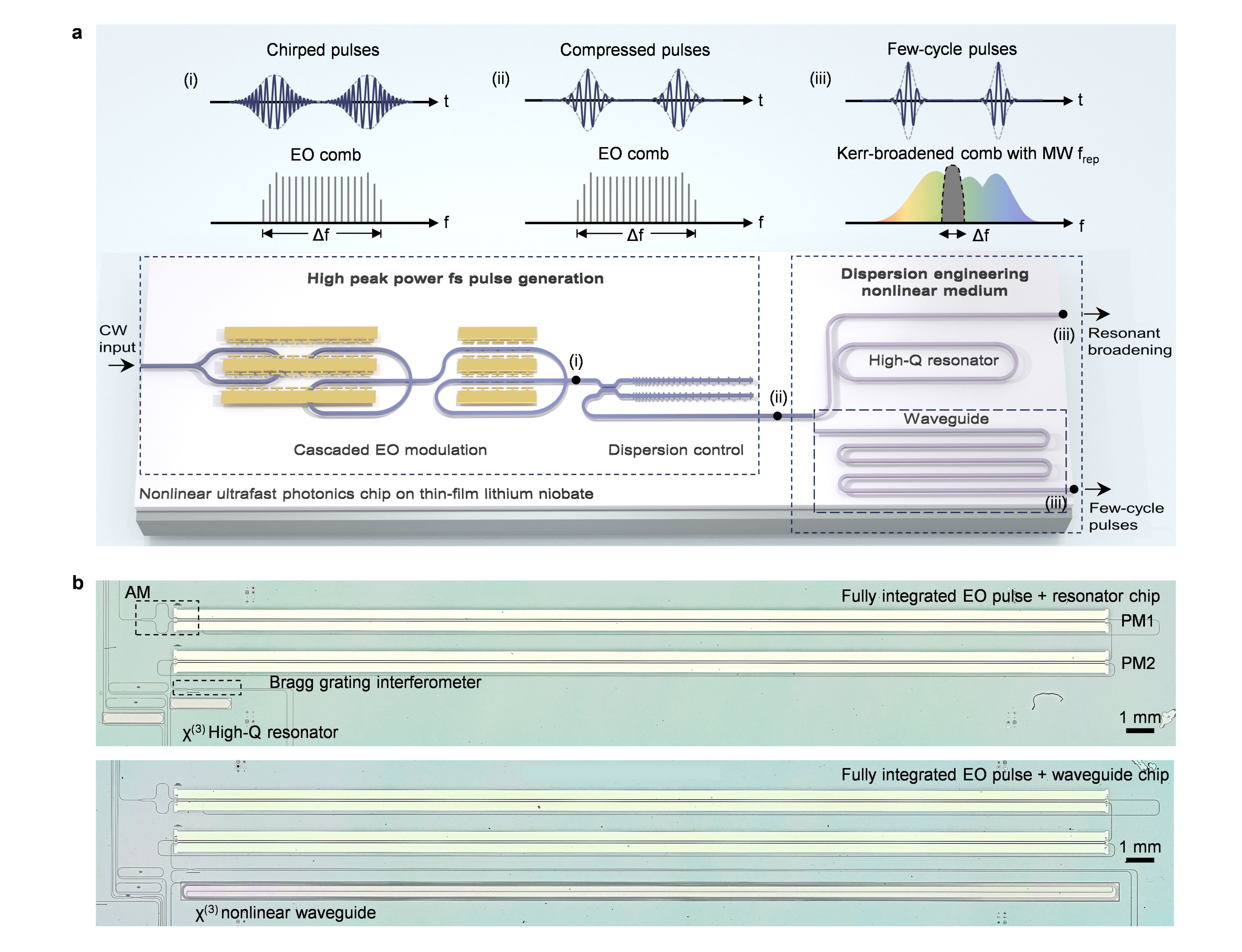}
    \caption{
    \textbf{Integrated architecture for ultrashort pulse and broadband comb generation.}
    \textbf{a}, Conceptual overview of the TFLN platform. A CW input is converted into (i) a broadband EO comb via cascaded amplitude and phase modulation, (ii) compressed into Fourier-transform limited femtosecond pulses by integrated dispersion control, and (iii) injected into a dispersion-engineered nonlinear stage. The resulting high-peak-power pulses drive Kerr nonlinear dynamics, enabling few-cycle compression and broadband spectral expansion in either a high-$Q$ resonator or a nonlinear waveguide. Top panels illustrate the corresponding evolution in the time and frequency domains.
    \textbf{b}, Optical micrograph of representative monolithic implementations. Top: fully integrated resonator-based platform combining EO modulation, on-chip dispersion control, and a high-$Q$ nonlinear resonator within a $37 \times 5$~mm footprint. Bottom: integration with a \SI{30}{cm} nonlinear waveguide. Both the nonlinear waveguide and resonator regions are air-cladded with a \SI{400}{nm} etch depth for anomalous dispersion engineering, whereas the EO modulation and linear dispersion elements are fully cladded with a \SI{300}{nm} etch depth for efficient EO mode overlap.
    }
    \label{fig1:concept}
\end{figure*}

Integrated nonlinear optics underpins a rapidly expanding range of applications, from precision metrology~\cite{Diddams_combreview_2020,Spencer_optical_synthesizer_2018}, ultrafast optics~\cite{Carlson_subcycleEO_2018}, microwave photonics~\cite{Feng2024Integrated,Qi_LNcomb_ranging_2025} to photonic computing~\cite{Xu2021TOPS} and quantum information processing~\cite{Moody2022Roadmap}. These complex photonic systems typically rely on diverse nonlinear light–matter interactions to control and transform electromagnetic fields across broad frequency, temporal, and energy scales. Therefore, scaling nonlinear photonic circuits requires more than achieving state-of-the-art performance in individual devices, but also multiple nonlinear and dispersive functionalities to operate together---ideally within a single material platform and on a single circuit---to minimize the additional loss and complexity associated with heterogeneous integration~\cite{Brodnik2026Monolithic}. In addition, nonlinear interactions depend strongly on the dispersion profile and nonlinearly on instantaneous optical power. Together, these present outstanding challenges for a system-level co-design of functionality, loss, and power flow, together with advanced fabrication capable of integrating disparate device geometries within a suitable single low-loss nonlinear optical platform.

Ultrafast photonics exemplifies this challenge. Conventional ultrafast sources rely on multi-stage bulk optical systems combining mode-locked pulse sources with a series of dispersive and nonlinear elements for spectral broadening and temporal compression~\cite{Popmintchev_HHG_2012,Piccoli_visiblepulse_2021}, while recent approaches explore replacing the gas-filled multi-pass cells and nonlinear fiber systems with integrated nonlinear structures \cite{Wu_UVcomb_2024,Ludwig_astrocomb_TFLN_2024,Gray_two_color_soliton_2026,Anderson_resonant_SCG_2021,Obrzud_microphotonic_astrocomb_2019,Choi_soliton_compression_2019, Oliver_soliton_compression_2021,Xu_pulsepumped_Kerr_2021,Li_pulsepumped_soliton_2022,Wei_KINQ_Bao_2026} which leverage the enhanced effective nonlinearity due to tight optical confinement. On the other hand, developing pulse sources on-chip is another active research direction, including dissipative Kerr soliton microcombs~\cite{Liu_microwave_microcomb_2020,Yi_soliton_2015, Helgason_soliton_efficiency_2023,Song_Kerrcomb_TFLN_2026}, heterogeneously integrated mode-locked lasers~\cite{Cuyvers_MLL_2021,Guo_MLL_LN_2023, Qiu_Mamyshev_MLL_2025}, and electro-optic (EO) frequency combs~\cite{Yu_femtosecondLN_2022,Cheng_single_drive_EO_comb_2024,Huang_TFLN_dispersion_compensator_2024}. However, unifying the ultrashort pulse generation with the subsequent nonlinear processes on a single material platform or photonic circuit has not been demonstrated. The challenges remain in not only the complexity of integrating multiple nonlinear and dispersive elements or even materials, but also simultaneously achieving Fourier-transform (FT)-limited femtosecond pulse duration, sufficient pulse energy, and peak power to directly drive subsequent nonlinear processes on chip.
Thin-film lithium niobate (TFLN) is particularly well suited to overcome these limitations as it simultaneously possesses a strong EO response, second-- and third--order optical nonlinearities, low optical loss and the tight optical confinement necessary for dispersion engineering~\cite{Boes2023Lithium,Zhu_TFLN_review_2021}.

Here we demonstrate a low-loss multi-nonlinear TFLN photonic circuit co-designed from continuous-wave (CW) excitation to EO ultrashort pulses to pulse-pumped nonlinear optics. As illustrated in~\cref{fig1:concept}a, CW light is (i) transformed through cascaded amplitude and phase modulation into a broadband EO frequency comb and a highly chirped pulse train at a microwave rate, (ii) linearly compressed through a dispersive grating interferometer to form high-peak-power femtosecond pulses, and (iii) delivered to a dispersion-engineered nonlinear waveguide or high-$Q$ resonator to drive pulse-pumped Kerr dynamics. 

Using an approximately \SI{20}{cm} integrated optical waveguide through the EO synthesis and compression stages, we generate \SI{230}{fs} FT-limited pulses at a \SI{30.7}{GHz} repetition rate, directly set by the microwave drive. Deliberately working at such high repetition rates $f_{\mathrm{rep}}$, where the pulse energy $E_{\mathrm{pulse}}$ for a given average optical power follows $P_{\mathrm{ave}}=f_{\mathrm{rep}}E_{\mathrm{pulse}}$, our system yields a pulse energy of \SI{3.3}{pJ} at an average power of 100 mW, and is compatible with interfacing to on-chip resonators for further resonant enhancement. This represents more than a twofold reduction in pulse duration and orders-of-magnitude increase in pulse energy compared to previous EO pulse synthesis~\cite{Yu_femtosecondLN_2022}. More importantly, we show that it provides sufficient peak power to directly drive subsequent nonlinear interactions in both non-resonant waveguides and resonant cavities on the same TFLN platform.

In a \SI{0.3}{m}-long dispersion-engineered waveguide, high-order soliton dynamics compress the pulses to \SI{35}{fs} (6.7 optical cycles), accompanied by coherent spectral broadening exceeding \SI{330}{nm} bandwidth and near 1,400 comb lines. Achieving few-cycle pulse synthesis directly from CW light using exclusively non-resonant photonic elements highlights the strength of light--matter interactions attainable in integrated nanophotonic circuits. In a complementary architecture, we demonstrate a fully monolithic implementation, as illustrated in~\cref{fig1:concept}b, integrating EO modulation, dispersive compression, and a million-$Q$ air-cladded nonlinear resonator on a single TFLN chip. Dialing the pulse repetition rate to match the free spectral range of the TFLN racetrack cavity enables resonantly enhanced pulse pumping and coherent spectral broadening at 400 fJ pulse energies.

All components are seamlessly integrated on a \SI{600}{nm}-thick X-cut TFLN platform which combines strong EO and Kerr nonlinearities, photonic-crystal bandgap and dispersion engineering, low-loss interferometric coupling, and low-loss optical propagation. The EO and Kerr stages require substantially different waveguide cross-sections to simultaneously optimize microwave–optical interaction and electrode performance for EO modulation, while achieving anomalous group velocity dispersion (GVD) and strong optical confinement for Kerr nonlinear dynamics. By accommodating these disparate geometries and high-performance microwave electrodes within a unified fabrication process, we realize the full system monolithically on a single TFLN chip. Together, these results extend integrated EO pulse synthesis to both traveling-wave and resonant pulse-pumped nonlinear optics, establishing a route towards multifunctional nonlinear photonic circuits and integrated few-cycle EO light sources.

\section{Results}\label{sec2}

\subsection{Electro-optic pulse compression with integrated dispersion}

\begin{figure*}[t]   
    \centering
    \includegraphics[width=\textwidth]
    {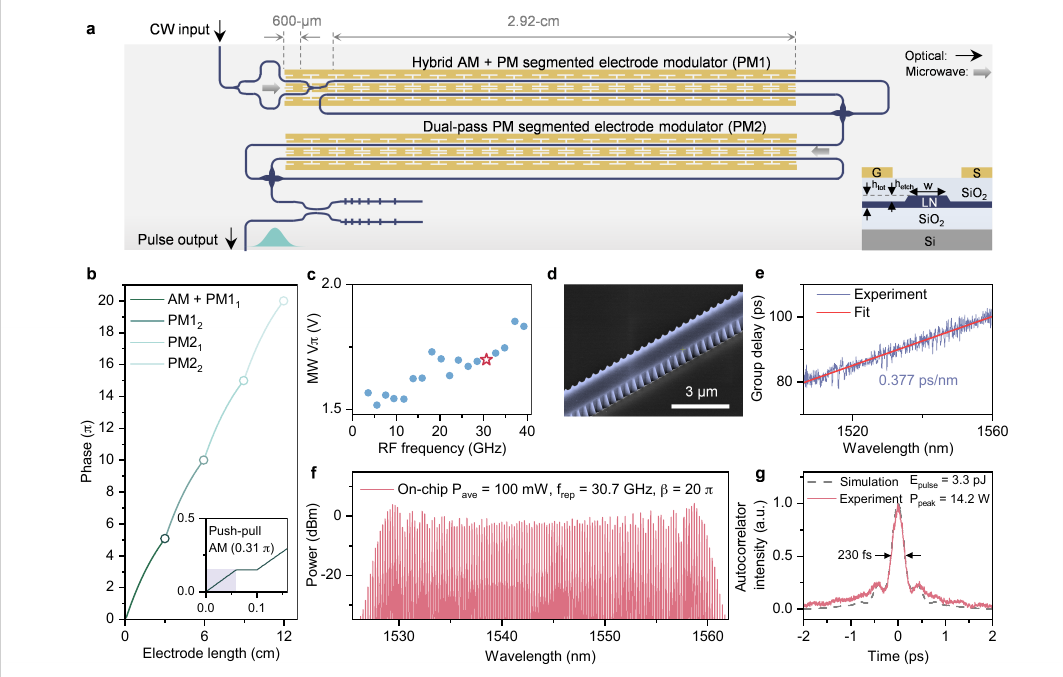}
    \caption{
    \textbf{Monolithic EO time-lens for femtosecond pulse synthesis.}
    \textbf{a}, Schematic of the hybrid AM-PM-PM with grating based dispersion architecture. A \SI{600}{\micro\meter} AM section is followed by two cascaded recycling PMs (PM1, PM2), with effective interaction lengths of \SI{5.84}{cm} and \SI{6.04}{cm}, respectively. The LN waveguide ($h_{\mathrm{tot}}=\SI{600}{nm}$, $h_{\mathrm{etch}}=\SI{300}{nm}$, $w=\SI{2}{\micro\meter}$) is embedded in a SiO$_2$ cladding on silicon. Two pairs of \SI{3}{cm}-long traveling-wave T-shaped electrodes with a \SI{3.7}{\micro\meter} signal–ground gap enable efficient EO modulation while maintaining low optical loss. The modulated light is dispersion-compensated using chirped Bragg gratings and extracted through a directional coupler.
    \textbf{b}, Calculated optical phase accumulation versus electrode length. The short AM provides a $0.31\pi$ phase bias, together with wavelength parking at a $-4$~dB transmission point ($\sim 0.57\pi$), yielding a total effective bias of $\sim 0.88\pi$ for high-extinction sinusoidal intensity carving. The cascaded PM1$_{1-2}$ and PM2$_{1-2}$ accomplish the quadratic phase accumulation of $20\pi$ .
    \textbf{c}, Measured microwave half-wave voltage ($V_{\pi}$) versus microwave frequency, with $V_{\pi}=\SI{1.7}{V}$ at \SI{30.7}{GHz} highlighted.
    \textbf{d}, SEM image of a chirped Bragg grating. Scale bar, \SI{3}{\micro\meter}.
    \textbf{e}, Measured (purple) and fitted (red) group delay dispersion of the chirped grating, yielding a slope of \SI{0.377}{ps/nm}, equivalent to \SI{21}{m} of standard SMF.
    \textbf{f}, EO frequency-comb spectrum at the output of the time-lens stage ($P_{\mathrm{ave}}=100$~mW, $f_{\mathrm{rep}}=\SI{30.7}{GHz}$, $\beta=20\pi$).
    \textbf{g}, Intensity autocorrelation trace of the compressed pulse train. The experimental trace (red) agrees well with simulation (dashed grey), resulting in \SI{230}{fs} pulses with \SI{3.3}{pJ} energy and \SI{14.2}{W} peak power.
}
    \label{fig2:fullyTL+G}
\end{figure*}
We realize the first stage of pulse synthesis using a monolithic EO time-lens chip, shown in~\cref{fig2:fullyTL+G}a, which coherently transfers microwave modulation to the optical field to synthesize broadband femtosecond pulses from a CW light. The architecture consists of a short amplitude modulator (AM), two cascaded recycling phase modulators (PMs), and a chirped Bragg grating interferometer for on-chip dispersion compensation. The device is fabricated on a \SI{600}{nm} X-cut TFLN platform with a \SI{300}{nm} etch depth and \SI{850}{nm} SiO$_2$ cladding (see Methods).
To reduce the circuit footprint, the \SI{600}{\micro\meter}-long AM shares the \SI{3}{cm}-long traveling-wave electrode with the first PM. Under the same microwave drive, the short AM provides a modulation depth of approximately $0.88\pi$, sufficient to carve a flat-top temporal aperture from the CW input, while the substantially longer interaction in the first PM imparts the strong phase modulation required for spectral broadening.
The two recycling PMs further enhance the EO interaction by allowing the optical field to traverse each approximately \SI{3}{cm}-long modulation section twice, providing effective modulation lengths of \SI{5.84}{cm} and \SI{6.04}{cm}, respectively. This extended interaction enables coherent accumulation of a modulation index of $20\pi$. Velocity matching between the optical and traveling microwave fields maintains efficient EO modulation over the full interaction length of 12 cm. As shown in~\cref{fig2:fullyTL+G}c, the measured $V_{\pi}$ is \SI{1.7}{V} at \SI{30.7}{ GHz} and remains between \SI{1.5}{V} and \SI{1.85}{V} from 3 to 40 {GHz} (see Methods).
Dispersion compensation and linear pulse compression is implemented using a symmetric Bragg-grating interferometer comprising a $2\times2$ directional coupler and two identical chirped gratings. Unlike off-chip circulator- or single-grating-based configurations that incur an intrinsic \SI{6}{dB} loss~\cite{Yu_femtosecondLN_2022}, the symmetric interferometer can in principle operate without intrinsic splitting loss for a 50:50 coupler. 
On this chip, the fabricated directional coupler exhibits a non-ideal $80{:}20$ power splitting ratio at \SI{1543}{nm}, resulting in an insertion loss of \SI{2.3}{dB} while the \SI{1.3}{mm}-long grating only contributes to \SI{0.1}{dB} of insertion loss (\SI{0.08}{dB/mm}). In a subsequent fabrication run, an optimized coupler reduced the total insertion loss of the complete grating interferometer to \SI{0.3}{dB} (see Table 1 in Methods), demonstrating the potential for ultralow-loss on-chip pulse compression.
The chirped gratings are formed by periodically modulating the waveguide width from \SI{0.8}{\micro\metre} to \SI{1.1}{\micro\metre}, as shown in the SEM image in~\cref{fig2:fullyTL+G}d. Linearly chirping the grating period from \SI{394}{nm} to \SI{410}{nm} over a length of \SI{1.3}{mm} produces a transmission bandwidth spanning \SIrange{1500}{1565}{nm} and an exact group-delay dispersion of \SI{0.377}{ps/nm} (\cref{fig2:fullyTL+G}e). This is needed to cover the optical bandwidth and quadratic spectral phase accumulated during EO modulation for the FT-limited pulse generation. This dispersion value is equivalent to approximately \SI{21}{m} of standard single-mode fiber.
At a repetition rate of \SI{30.7}{GHz}, we obtain a flat-top EO-comb spectrum with a \SI{10}{dB} optical bandwidth of \SI{31.5}{nm} centered at \SI{1543}{nm}, corresponding to an effective modulation index of $20\pi$ under a delivered microwave power of \SI{34.6}{dBm} per ground--signal--ground electrode (\cref{fig2:fullyTL+G}f). The pulse-synthesis circuit integrates approximately \SI{20}{cm} of optical waveguide, including \SI{12}{cm} of effective EO modulation length, with a total propagation loss of only \SI{0.7}{dB} (\SI{3.5}{dB/m}). This low-loss integration preserves an on-chip optical power of \SI{100}{mW}, corresponding to a pulse energy of \SI{3.3}{pJ}---nearly three orders of magnitude higher than previous integrated EO time lens pulse synthesis, where insertion losses approaching \SI{25}{dB} limited pulse energies to 1-2 femtojoule level~\cite{Yu_femtosecondLN_2022}. Following on-chip dispersion compensation, the pulses reach a FT-limited duration of \SI{230}{fs} and a peak power of \SI{14.2}{W} (\cref{fig2:fullyTL+G}g), in excellent agreement with numerical simulation. The combination of sub-\SI{250}{fs} duration, picojoule pulse energy and tens-of-watts peak power transforms the integrated EO time lens from a pulse-synthesis element into an ultrafast pump capable of directly driving subsequent nonlinear dynamics.

\subsection{Few-cycle pulse generation via soliton self-compression}
\begin{figure*}[t]   
    \centering
    \includegraphics[width=\textwidth]
    {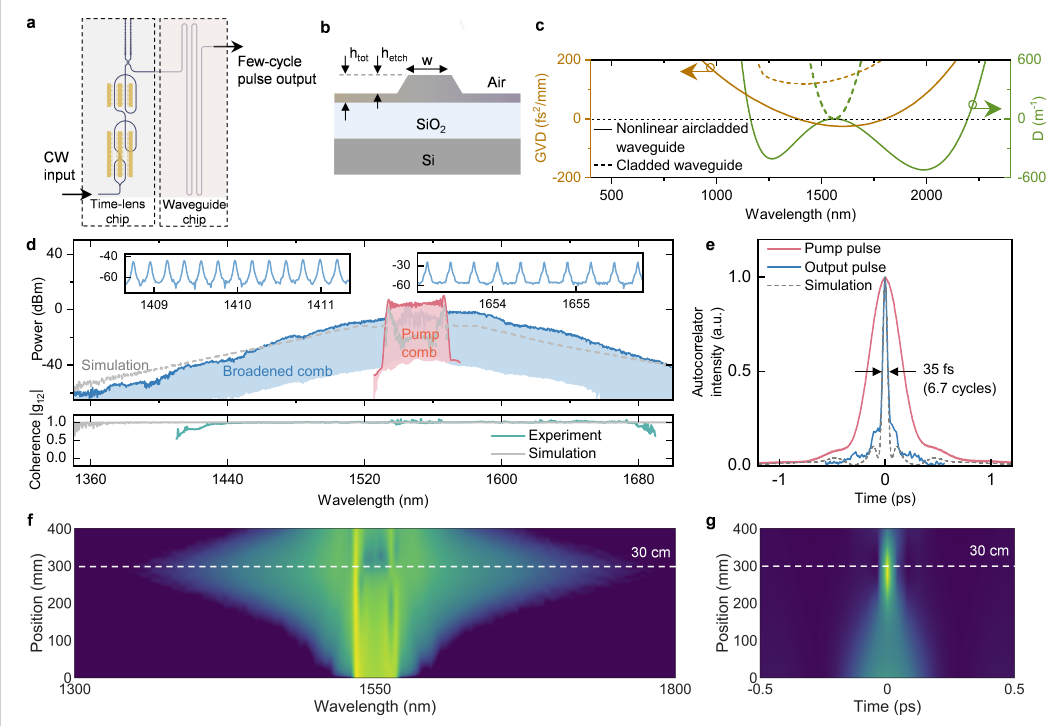}
    \caption{
    \textbf{Few-cycle pulse generation using a hybrid time-lens and dispersion-engineered nonlinear waveguide.}
    \textbf{a}, Hybrid configuration comprising an EO time-lens chip followed by a separate nonlinear lithium niobate waveguide for soliton-driven spectral broadening and temporal compression.
    \textbf{b}, Cross-sectional geometry of the air-cladded nonlinear LN waveguide with total thickness $h_{\mathrm{tot}}=\SI{600}{nm}$, etch depth $h_{\mathrm{etch}}=\SI{400}{nm}$, and width $w=\SI{1.6}{\micro\meter}$.
    \textbf{c}, Simulated group-velocity dispersion (GVD, left axis) and dispersion operator $D$ (right axis) for the SiO$_2$-cladded EO section (normal dispersion) and the air-cladded nonlinear waveguide (engineered anomalous dispersion of $\sim$\SI{-25}{fs^2/mm} near \SI{1550}{nm}), enabling high-order soliton dynamics.
    \textbf{d}, Measured pump comb (red) and spectrally broadened output (blue) after nonlinear propagation, in agreement with simulation (grey dashed). Insets show resolved comb lines at both spectral edges. The lower panel presents the measured (green) and simulated (gray) first-order spectral coherence across the broadened bandwidth.
    \textbf{e}, Autocorrelation traces of the pump (red) and compressed output pulses (blue), yielding \SI{35}{fs} (6.7 optical cycles) in agreement with simulation (grey dashed). The peak power of the output nonlinear compressed pulses reaches \SI{125}{W} while maintaining \SI{4.4}{pJ} pulse energy.
    \textbf{f,g}, Simulated spectral and temporal evolution along the nonlinear waveguide, showing maximal spectral broadening and shortest pulse duration near a propagation length of \SI{30}{cm}, with spectral coverage from \SI{1370}{nm} to \SI{1700}{nm} at the \SI{-60}{dB} level.
    }
    \label{fig3:waveguide}
\end{figure*}

To access the few-cycle regime, we interface the EO time-lens stage with a dispersion-engineered TFLN nonlinear waveguide, as illustrated in~\cref{fig3:waveguide}a.
The nonlinear waveguide is realized on the same \SI{600}{nm}-thick, X-cut TFLN platform as the EO circuit, but its cross-section and cladding are selectively modified to support strong Kerr nonlinear dynamics (\cref{fig3:waveguide}b). In \cref{fig1:concept}b, we showed that starting from the \SI{300}{nm} etch used for the EO section, an additional photolithography step and dry etching selectively deepens the nonlinear waveguide region by \SI{100}{nm}, producing a \SI{400}{nm} etch depth. The SiO$_2$ cladding is subsequently removed only over this region to form an air-cladded nonlinear waveguide, while the EO section remains oxide-cladded. The high-speed microwave electrodes are patterned by electron-beam lithography and deposited by metal evaporation as a final step (see Methods). This process enables substantially different dispersion-engineered waveguide geometries to be realized within the same TFLN chip without any heterogeneous optical material integration or lossy chip-to-chip coupling.
With a top width of \SI{1.6}{\micro\meter}, the resulting nonlinear waveguide exhibits anomalous GVD of approximately \SI{-25}{fs^2/mm} near \SI{1550}{nm} (\cref{fig3:waveguide}c), enabling high-order soliton formation and nonlinear self-compression. We note that beyond the local anomalous GVD at the pump wavelength, the dispersion operator $D$ reveals a broadband engineered higher order dispersion profile supporting octave-spanning spectrum via dispersive-wave generation (see Methods and Extended Data Fig.~3).
For sufficiently large soliton order ($N>1$), the balance between Kerr nonlinearity and anomalous GVD leads to periodic temporal breathing, with maximal compression occurring near the first oscillation~\cite{Oliver_soliton_compression_2021}.
The soliton order is given by $N^2 = L_D/L_{\mathrm{NL}}$, where $L_D = T_{\mathrm{in}}^2/|\beta_2|$ and $L_{\mathrm{NL}} = 1/(\gamma P_0)$ are the dispersion and nonlinear lengths, respectively, with $T_{\mathrm{in}}$ denoting the input-pulse duration. The optimal compression length can be estimated as $z_{\mathrm{opt}}/L_D = 0.32/N + 1.1/N^2$ ~\cite{Agrawal_NLFO_1989}. These relations directly connect the required interaction length to the input pulse duration, peak power, waveguide nonlinearity and dispersion, allowing the nonlinear stage to be co-designed with the output of the preceding EO pulse synthesizer.
For the present experiment, \SI{248}{fs} pulses generated by the EO time-lens stage are optically amplified and launched into a \SI{30}{cm}-long nonlinear waveguide with an on-chip peak power of \SI{43}{W}, corresponding to an average power of \SI{330}{mW} and a pulse energy of \SI{10.7}{pJ}. The measured propagation loss of the nonlinear waveguide is \SI{13}{dB/m}, corresponding to only \SI{3.9}{dB} over the full \SI{30}{cm} interaction length. We deliberately operate within this practical power range of the integrated EO source and exploit the extended low-loss nonlinear interaction to reach the first compression point. Waveguides ranging from \SIrange{10}{50}{cm} further verify the predicted length scaling (Extended Data Fig.~6).
Strong Kerr-driven spectral broadening is observed experimentally (\cref{fig3:waveguide}d), with the output spectrum extending from \SI{1370}{nm} to \SI{1700}{nm} at the \SI{-60}{dB} level. Individual comb lines remain clearly resolved at both spectral edges, demonstrating preservation of the \SI{30.7}{GHz} comb structure throughout the nonlinear broadening process.
To quantify the spectral coherence, we perform delayed self-interference measurements and extract the first-order degree of coherence $|g_{12}^{(1)}(\omega)|$ from the wavelength-dependent fringe visibility~\cite{Gaeta_chipcomb_2019} (see Methods). The measured coherence remains close to unity across the collected broadened spectrum.
At the output of the nonlinear waveguide, the pulses are compressed from \SI{248}{fs} to \SI{35}{fs}, corresponding to 6.7 optical cycles, while retaining a pulse energy of \SI{4.4}{pJ} and reaching a peak power of \SI{125}{W}. The measured pulse duration agrees closely with the simulated value of \SI{36}{fs} at the designed propagation length. Numerical modeling using the generalized nonlinear Schr\"odinger equation (NLSE), incorporating the simulated dispersion, measured nonlinear coefficient, and propagation loss, reproduces the observed spectral and temporal evolution (\cref{fig3:waveguide}f,g). Both the broadest spectrum and shortest pulse occur near \SI{30}{cm}, confirming operation near the designed first soliton-compression point.

These results demonstrate an all-traveling-wave route from CW excitation to few-cycle light, in which deterministic EO pulse synthesis directly drives soliton self-compression, which continues to increase the accessible on-chip peak power.

\begin{figure*}[t!]   
    \centering
    \includegraphics[width=\textwidth]
    {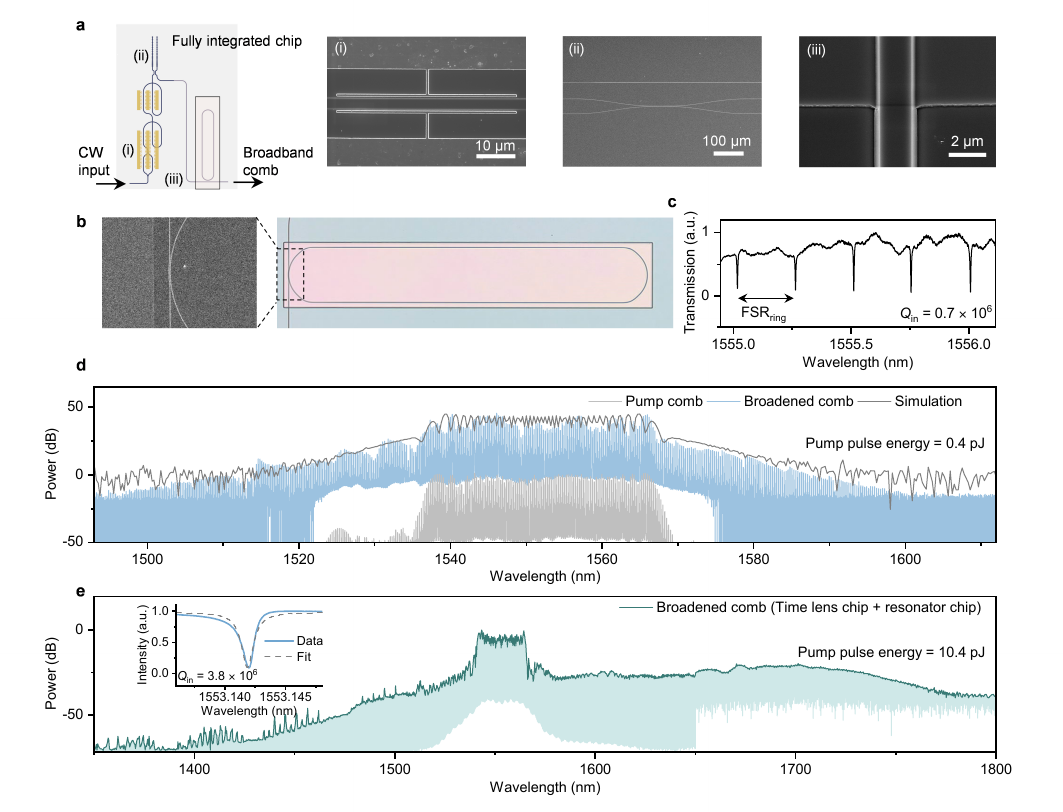}
    \caption{
    \textbf{Pump-pulse-driven nonlinear spectral broadening in a fully integrated resonator chip.}
    \textbf{a}, Schematic of the fully integrated chip architecture, in which the EO time-lens directly generates pump pulses that drive a nonlinear LN microring resonator to produce a broadband comb (the microscope image of this chip is shown in~\cref{fig1:concept}b). The resonator shares the same geometry in its straight waveguide part with the waveguide in~\cref{fig3:waveguide}, oriented perpendicular to the crystal $z$ axis. Panels (i)–(iii) show SEM images of the segmented electrode in the time-lens part, directional coupler, and the transition region from the time-lens section to the deeply etched nonlinear resonator ($\sim$400-nm etch depth), respectively.
    \textbf{b}, SEM image (left) and optical micrograph (right) of the dispersion-engineered nonlinear LN microring resonator integrated on the same chip.
    \textbf{c}, Measured transmission spectrum of the resonator, yielding an intrinsic quality factor of $Q_{\mathrm{in}} = 0.7\times10^{6}$. The resonator free spectral range ($\mathrm{FSR}_{\mathrm{ring}}$) is designed to match the microwave drive frequency of the time-lens.
    \textbf{d}, Optical spectra of the pump comb and the broadened comb obtained by directly injecting CW into the chip input, together with the simulated broadened spectrum (grey). The pulse energy right before the resonator is \SI{0.4}{pJ}.
    \textbf{e}, Spectral broadening obtained using a separate resonator chip. A broadband spectrum spanning \SI{1400}{nm} to beyond \SI{1800}{nm} at the \SI{-60}{dB} level is achieved, enabled by a higher pump pulse energy of \SI{10.4}{pJ}, with an intrinsic resonator quality factor of $3.8\times10^{6}$ (inset).
    }
    \label{fig4:resonator}
\end{figure*}

\subsection{Pulse-driven nonlinear broadening in a fully integrated resonator}
A distinct advantage of EO pulse synthesis is its flexible repetition rate in the microwave regime, which naturally matches the free spectral ranges (FSRs) of chip-scale optical resonators. At repetition rates of tens of gigahertz, successive pulses can be synchronized with compact high-$Q$ resonators, translating picojoule incident pulses to nanojoule intracavity energies while retaining tens-of-gigahertz repetition rates. This regime is difficult to access directly with conventional solid-state mode-locked lasers operating at megahertz repetition rates, which would require resonator round-trip lengths approaching the meter scale for synchronous pumping.
Leveraging this compatibility, we demonstrate the full integration of EO pulse synthesis, dispersion control and resonantly-enhanced $\chi^{(3)}$ nonlinear optics in a monolithic time--lens-resonator circuit, in which EO-synthesized femtosecond pulses directly drive a nonlinear microring resonator (\cref{fig4:resonator}a \& \cref{fig1:concept}b). The segmented-electrode-based EO modulators, directional coupler, dispersive element, and nonlinear microring are co-fabricated on the same \SI{600}{nm}-thick, X-cut TFLN substrate. The nonlinear resonator employs the same dispersion-engineered cross-section and fabrication process as the traveling-wave nonlinear waveguide in \cref{fig3:waveguide}, enabling both traveling-wave and resonant Kerr functionalities within the same integrated TFLN architecture. The SEM images in \cref{fig4:resonator}a highlight (i) the segmented-electrode region of the time-lens stage, (ii) the directional coupler and (iii) the transition into the deeply etched nonlinear resonator region. The performance and loss of the individual components are characterized independently (see Methods), allowing the pulse energy delivered to the resonator to be accurately determined.

The microresonator geometry is shown in \cref{fig4:resonator}b. The resonator operates near critical coupling and exhibits an intrinsic quality factor of $Q_{\mathrm{in}} = 0.7\times10^6$ on this chip (\cref{fig4:resonator}c) with its FSR designed to coarsely match the microwave repetition rate of the integrated time-lens source, such that successive EO-synthesized pulses arrive synchronously with the circulating intracavity field. The electronic tunability of the EO repetition rate further enables fine matching to the fabricated resonator FSR, providing a practical route to synchronous resonant pumping without requiring exact fabrication-level matching of the two frequencies.
We then inject a CW laser into the monolithic circuit. The integrated time-lens stage converts the CW field into \SI{386}{fs} pulses, which subsequently propagate directly into the nonlinear resonator without leaving the chip. After accounting for the calibrated loss through the EO and dispersive stages, the pulse energy delivered to the resonator is approximately \SI{0.4}{pJ}. The measured pump comb (light grey) and broadened output spectrum (blue) are shown in \cref{fig4:resonator}d. The measured spectral envelope is reproduced by numerical simulation (gray) using the time-lens-generated pulse as the driving field of the resonator. At this sub-picojoule-level pulse energy, resonant field enhancement provides sufficient nonlinear interaction to generate spectral broadening with more than 430 comb lines and beyond \SI{107}{nm} bandwidth.

To explore the performance accessible with increased pulse energy and higher resonator quality factor, we additionally study a two-chip configuration comprising a time-lens source followed by a separate high-$Q$ microring resonator with $Q_{\mathrm{in}}=3.8\times10^6$. Here, \SI{365}{fs} pulses are externally amplified before injection, providing an on-chip pulse energy of \SI{10.4}{pJ} (\SI{320}{mW} average power). The resulting spectrum broadens to more than \SI{400}{nm} and more than 1560 comb lines (\cref{fig4:resonator}e). Notably, we observe that the spectral broadening at this pulse-energy level is primarily dispersion limited, as a separate resonator with a lower intrinsic quality factor of 0.5 million produces a comparable spectral bandwidth (Extended Data Fig.~\ref{ext:RT500k}). Nevertheless, the monolithic architecture demonstrates resonant pulse pumping at sub-picojoule pulse energies within a fully-integrated TFLN circuit. 

Whereas the traveling-wave architecture exploits extended low-loss interaction length to access few-cycle pulses, the resonator architecture leverages coherent intracavity enhancement to strengthen nonlinear interactions at substantially lower pulse energies. These complementary approaches establish interaction length and resonant enhancement as two degrees of freedom of design for scaling pulse-driven nonlinear photonic circuits on TFLN.

\section{Conclusion and outlook}\label{sec3}
In summary, we establish TFLN as a unified multi-functional nonlinear platform in which deterministic EO femtosecond pulse synthesis interfaces directly with $\chi^{(3)}$ Kerr-mediated ultrafast nonlinear dynamics. By linking an on-chip time-lens source to dispersion-engineered waveguides and resonators, we realize a chip-scale pathway from CW input to few-cycle pulses and broadband coherent comb generation at microwave rates, complementary to 100-MHz-rate integrated mode-locked lasers, and offering the distinct advantages of direct electronic control and synchronization. Specifically, we generate a direct EO-transduction-based FT-limited \SI{230}{fs} pulse train at a repetition rate of \SI{30.7}{GHz} and a pulse energy of 3.3 pJ, followed by nonlinear compression to \SI{35}{fs} with coherent spectral broadening exceeding \SI{330}{nm}, and demonstrate synchronously pulse-driven resonant optical frequency comb generation, starting from a CW laser input on a fully monolithic TFLN time-lens resonator chip.

Further performance gains can be achieved by improving the EO pulse source itself. The present recycling phase modulators provide an effective interaction length of approximately \SI{12}{cm} and a modulation index of $20\pi$, while longer recycling modulators with effective interaction lengths reaching \SI{36}{cm} could further increase the modulation index and comb bandwidth under comparable microwave drive~\cite{Lee_flat_top_EO_comb_2025}, pointing to a potentially exciting performance breakthrough of direct EO pulse synthesis of \SI{120}{fs} for downstream nonlinear interactions. The wavelength flexibility of EO synthesis further expands the accessible nonlinear landscape: extending microwave-rate femtosecond pulse generation towards \SI{1}{\micro\meter} could, for example, directly pump octave-spanning supercontinuum generation and optical parametric oscillators operating at sub-pJ pulse energies~\cite{Sekine2025Multioctave,Jankowski2020Ultrabroadband}. In the present fully monolithic circuits, we observe component-level fabrication and loss variation across such a deep circuit. We therefore independently fabricate and characterize the EO modulators, dispersive gratings, waveguide transitions, and nonlinear elements, whose best measured performance corresponds to a projected total insertion loss of \SI{3}{dB} for the complete architecture (Methods and Table~\ref{tab:loss_budget}). Simulations based on these experimentally measured component losses indicate that, with less than \SI{1}{W} of on-chip CW input power, a fully integrated architecture can support pulse compression to \SI{18}{fs} and octave-spanning spectra in both traveling-wave and resonator configurations. Harnessing the strong $\chi^{(2)}$ nonlinearity and established domain-engineering capability of TFLN for frequency conversion could enable octave self-referencing and open a path toward chip-scale ultralow-noise microwave synthesis~\cite{Kudelin_microwave_2024,Sun_OFD_2024,Zhao_alloptical_2024,He_hQ_microwave_2024} —a long-standing goal at the intersection of optical atomic clocks, frequency metrology, and integrated photonics.

Beyond these immediate improvements, combining higher-peak-power EO pulses with higher-$Q$ resonators could substantially increase intracavity field strengths and provide access to regimes of extreme nonlinear light--matter interaction that remain difficult to reach in one integrated photonic circuit. More broadly, the ability to synthesize, compress, resonantly enhance, and nonlinearly transform ultrafast fields within a common material platform creates opportunities for cascaded nonlinear processes to be co-designed as a complete photonic system, rather than implemented as isolated nonlinear devices. The combination of femtosecond pulse duration, picojoule-level pulse energy, microwave-rate operation and direct electronic control could enable a broad range of integrated ultrafast photonic systems, spanning precision metrology, spectroscopy, waveform synthesis and quantum photonics.

\bibliography{main}

\section*{Methods}
\subsection*{Device fabrication}

The devices were fabricated on an X-cut lithium-niobate-on-insulator (LNOI) wafer consisting of a \SI{600}{nm} LN device layer on a \SI{2}{\micro m} buried oxide atop a silicon substrate. Waveguide and resonator structures were first defined by electron-beam lithography (EBL) using hydrogen silsesquioxane (HSQ) resist, followed by inductively coupled plasma (ICP) dry etching to form partially etched ridge waveguides.

The air-cladded nonlinear regions were subsequently patterned using photoresist by direct laser writing. A second dry-etch step was then performed to define the deeply etched nonlinear waveguides. Intermediate wet-etch processes using buffered oxide etchant (BOE), potassium hydroxide (KOH), and photoresist remover were used to remove residual layers and finalize the waveguide geometry with an etch depth of \SI{400}{nm}.

After fabrication of the nonlinear sections, a SiO$_2$ cladding layer was deposited over the entire chip by plasma-enhanced chemical vapor deposition (PECVD). Windows were then opened above the nonlinear regions by patterning photoresist with direct laser writing, followed by BOE etching of the SiO$_2$, leaving the nonlinear waveguides air-cladded.

For the oxide-cladded EO sections, microwave electrode patterns were defined in a second EBL step using poly(methyl methacrylate) (PMMA) resist. Metal electrodes were then deposited by evaporation and lift-off to complete the electro-optic modulators.

\subsection*{Optical loss characterization}

\begin{table*}[t]
\centering
\caption{\textbf{Loss budget of different integrated chip architectures.}}
\renewcommand{\arraystretch}{1.2}
\setlength{\tabcolsep}{6pt}
\resizebox{\textwidth}{!}{%
\begin{tabular}{lcccc}
\toprule

\textbf{Component} &
\textbf{\makecell{Integrated\\time-lens chip\\(dB)}} &
\textbf{\makecell{fully integrated\\time-lens-driven\\resonator chip (dB)}} &
\textbf{\makecell{fully integrated\\time-lens-driven\\waveguide chip (dB)}} &
\textbf{\makecell{Best-case\\component loss (dB)}} \\
\midrule

\textbf{20-cm EO modulators} & 0.7 & 12 & 1.7 & 0.7 \\
\textbf{Grating interferometer} & 2.3 & 2 & 0.3 & 0.3 \\
\textbf{Transition} & NA & 0 & \multirow{2}{*}{5} & 0 \\
\textbf{Air-cladding waveguide} & NA & NA &  & 2.1 (0.07 dB/cm) \\
\textbf{Air-cladding resonator ($Q$)}  & NA &  $0.7 \times 10^{6}$ (0.57 dB/cm) & NA & $5.5 \times 10^{6}$ (0.07 dB/cm)\\

\midrule
\textbf{Total} & 3 & 14 & 7 & $\sim 3$ \\

\bottomrule
\end{tabular}
}
\label{tab:loss_budget}
\end{table*}
The optical loss of each functional section was independently characterized to establish a quantitative loss budget for the different chip architectures. The results are summarized in Table~\ref{tab:loss_budget}. Because the nonlinear dynamics are governed primarily by the pulse peak power delivered to the nonlinear section, this loss analysis is essential for interpreting the performance of both the waveguide and resonator devices.

\begin{figure*}[t]   
    \centering
    \includegraphics[width=\textwidth]{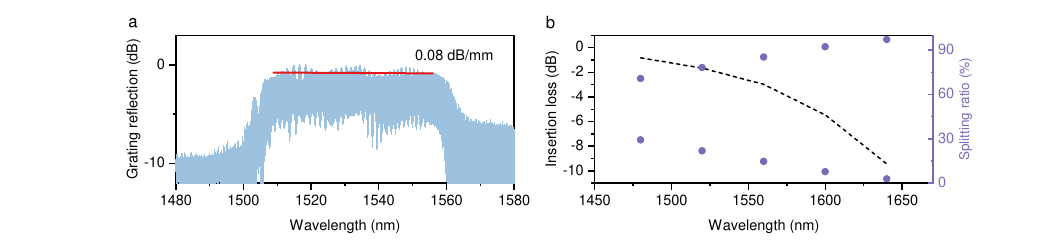}
    \extfigurecaption{
    \textbf{Grating coupler and interferometer loss characterization.}
    \textbf{a}, Measured reflection spectrum of the Bragg grating in time-lens stage after normalized by laser input power spectrum. The fitted spectral envelope indicates \SI{0.08}{dB/mm} transmission loss within the operating bandwidth.
    \textbf{b}, Wavelength-dependent insertion loss (dashed curve, left axis) of the grating interferometer and splitting ratio (purple circles, right axis) of the directional coupler.
}
    \label{ext:gratingloss}
\end{figure*}

For the integrated time-lens chip, the propagation loss of the EO modulator section was measured directly, whereas the insertion loss of the Bragg-grating interferometer was extracted from the measured splitting ratio together with spectral-envelope fitting, as shown in~\cref{ext:gratingloss}. The intrinsic transmission loss of the Bragg gratings was fitted to be \SI{0.08}{dB/mm}. These measurements yield a total on-chip loss of approximately \SI{3}{dB} for the present time-lens chip, while the propagation loss of the \SI{20}{cm} EO waveguide section alone is only \SI{0.7}{dB}. This low propagation loss is a key enabler of the high on-chip average power and pulse energy generated in the EO stage.

For the fully integrated resonator device, the transitions from the shallow-etched time-lens waveguide to the deeply etched nonlinear region introduces no measurable excess loss within experimental uncertainty, as verified by comparison with reference devices without the deeply etched transition region. The dominant pre-resonator loss instead originates from excess propagation loss in the monolithic time-lens section and the grating interferometer. Despite this loss penalty, intracavity enhancement provides sufficient peak power to observe substantial pulse-driven spectral broadening.

For the fully integrated waveguide device (\cref{fig1:concept}b), an additional loss of approximately \SI{7}{dB} is observed in the time-lens section after electrode fabrication. Comparison with separately fabricated
components indicates that this excess loss is associated with fabrication
variation in the monolithic process rather than the intrinsic propagation
loss of the EO waveguide.

To estimate the performance limit of the fully integrated architecture, we consider a projected scenario in which each section operates at its best measured loss: \SI{0.7}{dB} for the EO modulator section and \SI{0.3}{dB} for the Bragg-grating interferometer. Under these conditions, simulations predict substantially improved nonlinear performance for both monolithic implementations. For the waveguide device, the projected result is shown in~\cref{ext:bestwg}, where an on-chip CW input power of \SI{29}{dBm} and a pump pulse duration of \SI{208}{fs} yield pulse compression to \SI{18}{fs} together with octave-spanning spectral broadening. The corresponding projection for the resonator architecture is shown in~\cref{ext:bestring}, indicating octave-spanning comb generation under the same best-case loss assumptions. 

\subsection*{Microwave characterization}
\begin{figure*}[t]
    \centering
    \includegraphics[width=\textwidth]{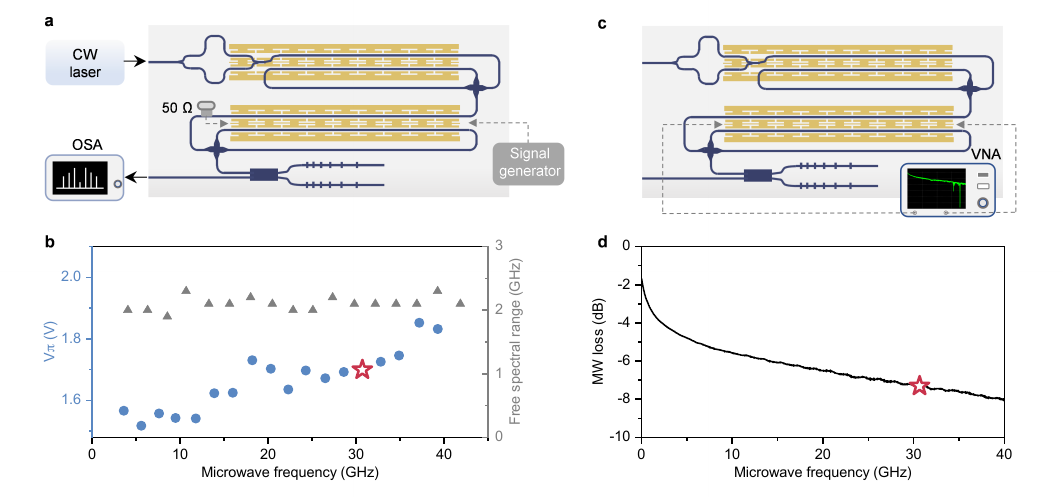}
  \extfigurecaption{
  \textbf{Microwave characterization of the electro-optic modulators.}
  \textbf{a}, Experimental setup for measuring the microwave half-wave voltage ($V_{\pi}$) using the pump-depletion method~\cite{Yu_femtosecondLN_2022}. The phase of a CW optical signal is modulated by the \SI{3.02}{cm} PM$_2$, and the modulation index and corresponding $V_{\pi}$ are extracted by fitting the optical spectrum measured on an optical spectrum analyzer (OSA).
  \textbf{b}, Extracted microwave $V_{\pi}$ (blue circles, left axis) and free spectral range of the modulation response (grey triangles, right axis) versus microwave frequency. The red star marks the operating frequency of \SI{30.7}{GHz}.
  \textbf{c,d}, Measurement setup and corresponding microwave transmission loss of the \SI{3.02}{cm} traveling-wave electrode, characterized using a vector network analyzer (VNA). The red star indicates the loss of \SI{7.33}{dB} at operating frequency of \SI{30.7}{GHz}.
}
\label{ext:RFVpi_MWloss}
\end{figure*}

The microwave response of the electro-optic modulators was characterized using pump-depletion measurements and vector network analysis.
The half-wave voltage ($V_{\pi}$) and the FSR of the modulation response were measured using the pump-depletion method with the experimental setup shown in~\cref{ext:RFVpi_MWloss}a. The extracted $V_{\pi}$ values and the corresponding FSR of the modulation efficiency are plotted in~\cref{ext:RFVpi_MWloss}b.
The microwave transmission loss of the traveling-wave electrodes was measured using a vector network analyzer (VNA). After calibrating out the cable and probe losses, the microwave transmission of a \SI{3.02}{cm}-long electrode was measured, as shown in~\cref{ext:RFVpi_MWloss}c, d. The microwave insertion loss at the operating frequency of \SI{30.7}{GHz} is \SI{7.3}{dB}.

\subsection*{Pulse-duration characterization}

Pulse durations are determined from intensity autocorrelation measurements. Because the EO-synthesized and nonlinearly compressed temporal waveforms are not assumed to follow an exact Gaussian or sech$^{2}$ profile, the autocorrelation deconvolution factor is obtained from the corresponding simulated temporal waveform. Specifically, the simulated intensity profile $I(t)$ is numerically autocorrelated according to
\begin{equation}
G^{(2)}(\tau)
=
\int I(t)I(t-\tau)\,\mathrm{d}t,
\end{equation}
and the ratio between the intensity full width at half maximum and the autocorrelation full width at half maximum is used to convert the measured autocorrelation width into the reported pulse duration. The same procedure is applied to both the input and nonlinearly compressed output pulses.

\subsection*{Spectral coherence measurement}

The first-order spectral coherence of the broadened comb is characterized using a delayed self-interference measurement. The output light is divided into two optical paths and recombined after introducing a relative delay, producing wavelength-dependent spectral interference fringes that are recorded using an optical spectrum analyzer. The spectra from the two individual interferometer arms, $I_1(\omega)$ and $I_2(\omega)$, are also recorded separately to account for unequal optical powers in the two paths.

The spectral fringe visibility is determined from the upper and lower
envelopes of the interference spectrum according to
\begin{equation}
V(\omega)
=
\frac{I_{\max}(\omega)-I_{\min}(\omega)}
{I_{\max}(\omega)+I_{\min}(\omega)}.
\end{equation}
The magnitude of the first-order degree of coherence is then obtained as
\begin{equation}
\left|g_{12}^{(1)}(\omega)\right|
=
V(\omega)
\frac{I_1(\omega)+I_2(\omega)}
{2\sqrt{I_1(\omega)I_2(\omega)}}.
\end{equation}
For equal optical powers in the two interferometer arms, this expression
reduces to $|g_{12}^{(1)}|=V$. 

The numerical spectral coherence is evaluated using an ensemble of nonlinear
propagation simulations with independently seeded quantum noise. For each
realization, quantum noise is added to the input field before propagation
through the nonlinear waveguide. The first-order spectral coherence between
independent realizations is calculated as
\begin{equation}
\left|g_{12}^{(1)}(\omega)\right|
=
\frac{
\left|
\left\langle
E_i^{*}(\omega)E_j(\omega)
\right\rangle_{i\neq j}
\right|
}{
\sqrt{
\left\langle |E_i(\omega)|^{2}\right\rangle
\left\langle |E_j(\omega)|^{2}\right\rangle
}
},
\end{equation}
where the brackets denote averaging over independent noise realizations.
The extracted coherence therefore quantifies the preservation of deterministic spectral phase relationships across the
nonlinearly broadened comb.

\subsection*{Numerical simulation}

\subsubsection*{Time-lens stage}

The time-lens output is modeled by cascading amplitude modulation, phase modulation and dispersive compression. The optical field after the two recycling phase modulators is written as
\begin{equation}
E_{\mathrm{TL}}(t)
=
E_{\mathrm{AM}}(t)
\exp\!\left[i\phi_{1}(t)\right]
\exp\!\left[i\phi_{2}(t)\right],
\end{equation}
where $E_{\mathrm{AM}}(t)$ is the field after the amplitude modulator and
$\phi_{1,2}(t)$ are the phase shifts applied by the two phase modulators.
Dispersion compensation is modeled in the frequency domain as
\begin{equation}
\tilde{E}_{\mathrm{out}}(\omega)
=
\tilde{E}_{\mathrm{TL}}(\omega)
\exp\!\left[-\frac{i}{2}\beta_{2}L\omega^{2}\right].
\end{equation}
This model follows the electro-optic time-lens framework of Ref.~\cite{Yu_femtosecondLN_2022}.

\subsubsection*{Time-lens--driven waveguide}

Nonlinear pulse propagation in the air-cladded waveguide is modeled using the generalized nonlinear Schr\"odinger equation (NLSE),
\begin{equation}
\frac{\partial A}{\partial z}
=
-\frac{\alpha}{2}A
+i\hat{D}A
+i\gamma |A|^{2}A ,
\end{equation}
where $\alpha$ is the measured propagation loss, $\hat{D}$ is the dispersion operator and $\gamma$ is the effective Kerr nonlinear coefficient. In the simulations, the dispersion operator is constructed directly from the finite-element-calculated propagation constant,
\begin{equation}
\hat{D}(\omega)
=
\beta(\omega)-\beta(\omega_{0})
-\beta_{1}(\omega_{0})(\omega-\omega_{0}).
\end{equation}
The nonlinear coefficient is evaluated as
\begin{equation}
\gamma
=
\frac{n_{2}\omega_{0}}{cA_{\mathrm{eff}}}\eta_{\mathrm{NL}},
\end{equation}
with $n_{2}=1.8\times10^{-19}~\mathrm{m^{2}/W}$.

The input field is taken directly from the simulated time-lens output and propagated using an adaptive split-step Fourier method. Self-steepening and measured propagation loss are included.

\begin{figure*}[t]   
    \centering
    \includegraphics[width=\textwidth]{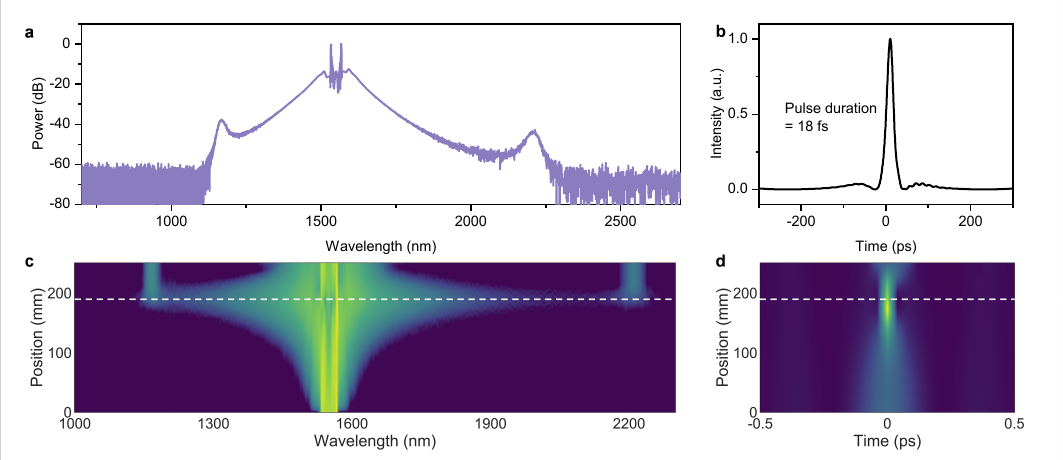}
    \extfigurecaption{
\textbf{Simulated pulse compression and spectral broadening for an optimized fully integrated time-lens--waveguide system.}
\textbf{a}, Simulated optical spectrum at the optimal compression point, $z=\SI{19}{cm}$, in a \SI{25}{cm}-long air-cladded TFLN waveguide, assuming the best measured loss for each component in the fully integrated architecture. The simulation assumes \SI{33}{dBm} off-chip pump power, corresponding to \SI{29}{dBm} on-chip continuous-wave power after \SI{4}{dB} facet-coupling loss. Additional losses include \SI{0.7}{dB} from the electro-optic modulator section, \SI{0.3}{dB} from the Bragg-grating interferometer, and a \SI{3}{dB} reduction from the amplitude modulator operated at the quadrature point, resulting in a pulse energy of \SI{10.3}{pJ} immediately before the nonlinear waveguide section. The resulting spectrum spans more than one octave at the \SI{-60}{dB} level.
\textbf{b}, Simulated temporal profile at $z=\SI{19}{cm}$, showing a minimum pulse duration of approximately \SI{18}{fs}.
\textbf{c,d}, Simulated spectral and temporal evolution, respectively, over the full \SI{25}{cm} waveguide. The dashed line marks the optimal compression position at $z=\SI{19}{cm}$ corresponding to \textbf{a,b}.
}
    \label{ext:bestwg}
\end{figure*}

For the optimized fully integrated waveguide architecture shown in
\cref{ext:bestwg}, the simulation uses the best measured component losses
listed in Table~\ref{tab:loss_budget}, corresponding to \SI{208}{fs},
\SI{10.3}{pJ} pulses immediately before the nonlinear waveguide. The pulse
reaches a minimum duration of approximately \SI{18}{fs} near
$z=\SI{19}{cm}$, accompanied by octave-spanning spectral broadening.

\subsubsection*{Time-lens--driven resonator}

Pulse-driven spectral broadening in the resonator is modeled using a generalized mean-field cavity equation, in which the intracavity field is driven by the pulsed output of the EO time-lens stage. Using the same dispersion operator and Kerr coefficient defined above, the slow-time evolution of the intracavity field envelope $A(t,\tau)$ is written as
\begin{equation}
\frac{\partial A}{\partial t}
=
\mathrm{FSR}
\left[
\mathcal{L}A
+i\gamma L |A|^{2}A
-\sqrt{\theta}\,A_{\mathrm{in}}(\tau)
\right],
\end{equation}
where the linear cavity operator is defined as
\begin{equation}
\mathcal{L}
=
-\frac{\alpha}{2}
-i\delta_{0}
+iL\hat{D}
+\sqrt{1-\theta}-1 .
\end{equation}
Here $\alpha$ is the intrinsic round-trip loss, $\delta_{0}$ is the pump--cavity detuning, $\theta$ is the power coupling coefficient, $L$ is the resonator round-trip length, and $A_{\mathrm{in}}(\tau)$ is the pulsed driving field obtained from the simulated time-lens output. The resonator free spectral range is matched to the modulation repetition rate when comparing with experiment.

The intrinsic and loaded quality factors ($Q_{\mathrm{in}}$ and $Q_{\mathrm{L}}$) used in the simulation are taken from the measured cavity transmission.
For the fully integrated resonator in~\cref{fig4:resonator}, we use $Q_{\mathrm{in}}=7.0\times10^{5}$ and $Q_{\mathrm{L}}=4.3\times10^{5}$, from which the corresponding cavity loss and coupling are determined.

The cavity dynamics are solved using the same adaptive split-step Fourier method as in the waveguide model. The linear cavity response is evaluated in the frequency domain and the Kerr nonlinear phase accumulation in the fast-time domain. Self-steepening is included, whereas Raman effects are neglected.

The through-port field is calculated as
\begin{equation}
A_{\mathrm{out}}(\tau)
=
\sqrt{\theta}\,A(\tau)
+
\sqrt{1-\theta}\,A_{\mathrm{in}}(\tau),
\end{equation}
from which the simulated output spectra are obtained.

The measured spectrum from the fully integrated time-lens--resonator device is compared with numerical simulation in~\cref{fig4:resonator}d. For this simulation, the driving field $A_{\mathrm{in}}(\tau)$ is taken from the simulated on-chip output of the monolithic time-lens stage and normalized to the experimentally determined pulse energy of approximately \SI{0.4}{pJ} immediately before the resonator. The calculated spectrum reproduces the measured spectral envelope.

\begin{figure*}[t]
    \centering
    \includegraphics[width=\textwidth]{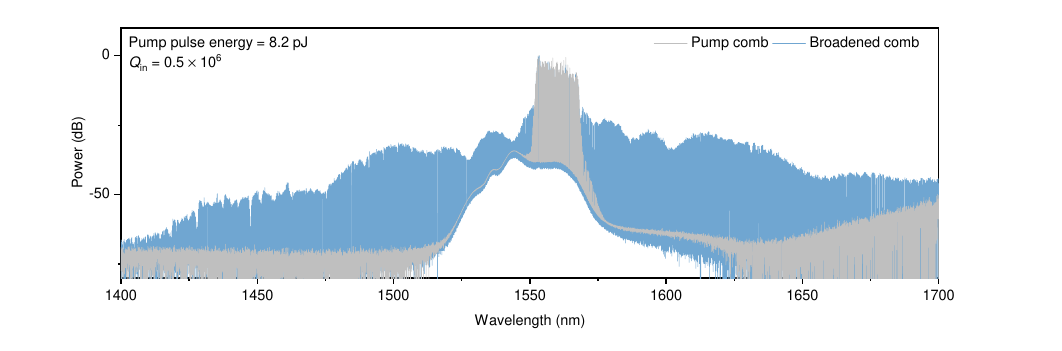}
     \extfigurecaption{
\textbf{Pulse-driven spectral broadening in a lower-$Q$ nonlinear resonator.}
Measured pump comb (grey) and spectrally broadened output comb (blue) from a separate-chip configuration comprising a time-lens source followed by a nonlinear TFLN resonator. The resonator has an intrinsic quality factor of $Q_{\mathrm{in}}=0.5\times10^{6}$ near critical coupling and is driven with an on-chip pump pulse energy of \SI{8.2}{pJ}. Despite its substantially lower intrinsic quality factor than the $Q_{\mathrm{in}}=3.8\times10^{6}$ resonator shown in Fig.~4e, a comparable spectral bandwidth is obtained, suggesting that in this pulse-energy regime the attainable bandwidth is more strongly constrained by resonator dispersion than by the intrinsic quality factor alone.
}
    \label{ext:RT500k}
\end{figure*}

To examine the role of resonator quality factor at higher pulse energies, we additionally characterize a separate resonator with $Q_{\mathrm{in}}=0.5\times10^{6}$. When driven with an on-chip pulse energy of \SI{8.2}{pJ}, this device produces a spectral bandwidth comparable to that obtained from the $Q_{\mathrm{in}}=3.8\times10^{6}$ resonator in~\cref{fig4:resonator}e (Extended Data Fig.~\ref{ext:RT500k}), suggesting that in this regime the attainable bandwidth is more strongly constrained by resonator dispersion than by the intrinsic quality factor alone.

\begin{figure*}[t]
    \centering
    \includegraphics[width=\textwidth]{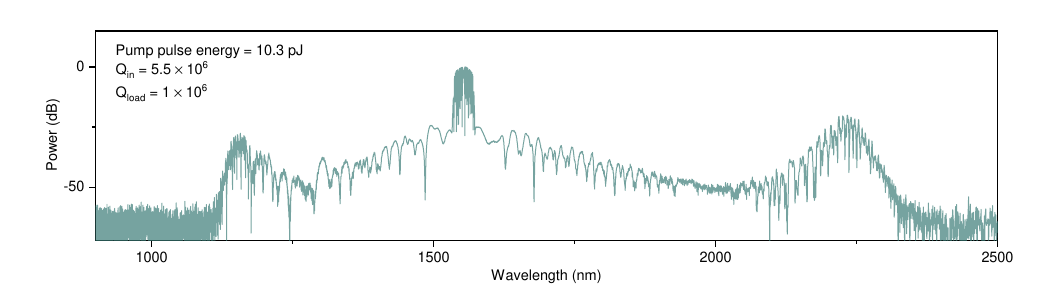}
    \extfigurecaption{
\textbf{Projected octave-spanning spectrum for an optimized fully integrated time-lens--resonator architecture.}
Simulated broadened comb spectrum assuming the best measured loss for each device section in the fully integrated architecture. The simulation uses \SI{208}{fs} pump pulses with an energy of \SI{10.3}{pJ} immediately before the nonlinear resonator. These pulses correspond to an off-chip continuous-wave input power of \SI{33}{dBm}, or \SI{29}{dBm} on chip after \SI{4}{dB} facet-coupling loss, together with \SI{0.7}{dB} loss through the electro-optic modulator section, \SI{0.3}{dB} insertion loss through the Bragg-grating interferometer, and a \SI{3}{dB} power reduction from the amplitude modulator operated at the quadrature point. Under these conditions, the nonlinear resonator is predicted to generate an octave-spanning spectrum extending from approximately \SI{1100}{nm} to \SI{2340}{nm}.
}
    \label{ext:bestring}
\end{figure*}

To estimate the performance accessible in an optimized fully integrated architecture, we further simulate the resonator using the best measured loss for each preceding device section. Under these conditions, the time-lens delivers \SI{208}{fs} pulses with an energy of \SI{10.3}{pJ} immediately before the resonator, producing an octave-spanning simulated comb extending from approximately \SI{1100}{nm} to \SI{2340}{nm} (\cref{ext:bestring}).

\subsection*{Pulse-driven propagation in waveguides of different lengths}
To examine the influence of nonlinear propagation length on pulse compression and spectral broadening, experiments were performed using air-cladded TFLN waveguides with lengths of \SI{50}{cm} and \SI{30}{cm}. In both cases, the waveguides were driven by the pulsed output of the EO time-lens stage described in the main text, and the output spectra and temporal profiles were compared with numerical simulations based on the generalized NLSE.

\begin{figure*}[t]   
    \centering
    \includegraphics[width=\textwidth]{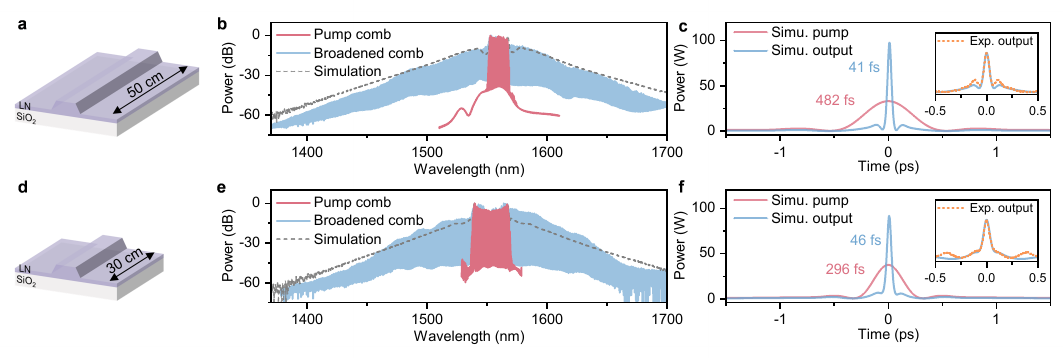}
   \extfigurecaption{
\textbf{Pulse compression and spectral evolution in waveguides of different lengths.}
\textbf{a}, Device schematic for a \SI{50}{cm} air-cladded TFLN waveguide used for pulse compression experiments.
\textbf{b}, Measured pump spectrum (red), broadened output spectrum (blue), and simulated spectrum (dashed line) for the \SI{50}{cm} waveguide.
\textbf{c}, Simulated temporal evolution showing compression of the input pulse from \SI{482}{fs} to \SI{41}{fs}. The inset shows the experimentally measured output autocorrelation trace.
\textbf{d}, Device schematic for a \SI{30}{cm} waveguide.
\textbf{e}, Measured pump and broadened output spectra together with the simulated spectrum for the \SI{30}{cm} waveguide.
\textbf{f}, Simulated temporal evolution showing compression of the input pulse from \SI{296}{fs} to \SI{46}{fs}, with the measured output autocorrelation trace shown in the inset.
}
    \label{ext:5030cm}
\end{figure*}

As shown in Extended Data Fig.~\ref{ext:5030cm}, both waveguide lengths exhibit strong spectral broadening and substantial pulse compression. For the \SI{50}{cm} waveguide, an input pulse of \SI{482}{fs} is compressed to approximately \SI{41}{fs}. For the \SI{30}{cm} waveguide, an input pulse of \SI{296}{fs} is compressed to approximately \SI{46}{fs}. The measured spectra and temporal profiles are in good agreement with the simulated spectral and temporal evolution along the waveguide.

\section*{Data availability}
The datasets generated and analyzed in the current study are available
from the corresponding authors on reasonable request.
\section*{Acknowledgements}
This work is supported by the DARPA Young Faculty Award (D23AP00252-02).
Device fabrication was performed at the John O’Brien Nanofabrication Laboratory at University of Southern California, the Nanolab at the University of California, Los Angeles, and Marvell Nanofabrication Laboratory at University of California, Berkeley.
M.Y. and Y.Y. are supported by the U.S. Department of Energy, Office of Science, Basic Energy Sciences, Materials Sciences and Engineering Division under Contract No. DE-AC02-05CH11231 within the Quantum Coherent Systems Program KCAS26.
The views, opinions and/or findings expressed are those of the authors and should not be interpreted as representing the official views or policies of the Department of Defense or the U.S. Government.

\section*{Author contributions}
M.Y. conceived the idea. 
C.-H.L. designed the chip with the help of X.R., C.C. and R.K.. 
C.-H.L. fabricated the devices and developed the fabrication processes with the help of C.C. and R.K..
X.R. carried out the experiments and analyzed the data with help from I.C. and L.Z..
X.R. performed the numerical simulation of time-lens stage and time-lens–driven resonator/waveguide.
X.R. performed the dispersion simulation with the help of Y.Y..
X.R., C.-H.L. wrote the manuscript with contribution from all authors. 
M.Y. and Z.C. supervised the project.

\section*{Competing interests}
C.-H.L., L.Z., Z.C. and M.Y. are involved in developing lithium niobate technologies at Opticore Inc.



\end{document}